\documentclass{desyproc}

\begin{document}
\title{Investigation of the rotation effects on high--density matter in hybrid stars}

\author{{\slshape Tomoki Endo}\\[1ex]
Division of Physics, Department of General Education, National Institute of Technology, Kagawa College, 355 Chokushi-cho, Takamatu, Kagawa 761-8058, Japan\\}

\contribID{98}

\confID{8648}  
\desyproc{DESY-PROC-2014-04}
\acronym{PANIC14} 
\doi  

\maketitle

\begin{abstract}
The equation of state (EOS) of high-density matter is still not clear and several recent observations indicate restrictions to EOSs. Theoretical studies should thus elucidate EOSs at high density and/or high temperature. Many theoretical studies have attempted to account for the effect of rotation of rapidly rotating neutron stars (pulsars), which are commonly observed astronomical objects having high-density interiors. Furthermore, neutron stars generate a strong magnetic field. Several recent studies indicate that this magnetic field exerts some restrictions on the EOS. Theoretical studies should thus incorporate these effects.
In this paper, we focus on the effect of rotation. We find that one of our EOSs is consistent with these observations, and another is inconsistent. We also find an important relation between radius and rotation.
\end{abstract}

\section{Introduction}
It is widely believed that quark matter exists in
high-temperature and/or high-density environments such as those of relativistic
heavy-ion collisions \cite{rhic} or the cores of neutron stars \cite{mad3,chen},
and
the ``deconfinement transition'' has been actively searched.
Theoretical studies using model calculations or based on the first
principle, lattice QCD \cite{rev}
have been also carried out by many authors to find the critical temperature of the deconfinement transition. 
Although many exciting results have been reported, the deconfinement transition
is not yet clearly understood. Many theoretical studies have suggested that the deconfinement 
transition is of first order in high-density cold matter \cite{pisa,latt}.  
 We thus assume that it is a first-order phase transition in the present work.
 We have given the equation of state (EOS) for the hadron--quark mixed phase taking
 into account the charge screening effect \cite{end2} without making 
 any approximations. We have investigated the inner structures of neutron stars as environments of quark matter \cite{end3,end4}.
Recently, many theoretical studies have attempted to account for the effect of the rotation of neutron stars \cite{kurk, orsa, webe, belv}. The results suggest that observations restrict the EOSs of theoretical calculations. Other studies have given the effect of the magnetic field \cite{dexh,chir,agui,mall} and it would thus be interesting to account for the magnetic effect in our EOS. However, as a first step, we focus on the effect of rotation.
 We thus apply our EOS to a stationary rotating star in this paper.

\section{Formalism and Numerical Results}
Our formulation was presented in detail in Ref.\ \cite{end2,end3} and is only briefly explained here.
The quark phase consists of {\it u}, {\it d}, and {\it s} quarks and
the electron. We incorporate the MIT bag model and assume a sharp boundary at the
hadron--quark interface. 
{\it u} and {\it d}
quarks are treated as massless and {\it s} as having mass
($m_s=150$MeV), and the quarks interact with each other via a
one-gluon-exchange interaction inside the bag.
The hadron phase consists of the proton, neutron and electron. The
effective potential is used to describe the interaction
between nucleons and to reproduce the saturation properties of nuclear matter.
In treating the phase transition, we have to consider the thermodynamic potential.
The total thermodynamic potential ($\Omega_\mathrm{total}$) 
consists of hadron, quark and electron contributions and the
surface contribution:
\begin{equation}
\Omega_\mathrm{total} = \Omega_\mathrm{H} +\Omega_\mathrm{Q}
 +\Omega_\mathrm{S}, 
\label{ometot}
\end{equation}
where $\Omega_\mathrm{H(Q)}$ denotes the contribution of the hadron (quark)
phase. We here introduce the surface contribution $\Omega_\mathrm{S}$,
parameterized  by the surface tension parameter $\sigma$, $\Omega_\mathrm{S} = \sigma S$, with 
$S$ being the area of the interface.
Note that $\Omega_\mathrm{S}$ may be closely related with the confining mechanism and unfortunately 
we have no definite idea about how to incorporate it. 
Many authors
have treated its strength as a free parameter and investigated 
how its value affects results \cite{pet,gle2,alf2}.
 We take the same approach in this study.
To determine the charge screening effect, we also make calculations 
without the screening effect \cite{end2, maru, maru2}. 
We then apply the EOS derived in our paper \cite{end2} to
the Tolman--Oppenheimer--Volkoff equation \cite{end3,end4}. 
We finally apply our EOS to a stationary rotating star.
However, it is difficult to consider the rotation effect in general relativity. We therefore make assumptions of 1) stationary rigid rotation (``uniform rotation''), 2) axial symmetry with respect to the spin axis; and 3) the matter being a perfect fluid.
Stationary rotation in general relativity has been reviewed in \cite{ster} and \cite{kurk}; we follow their calculation. We then apply our EOS to a stationary rotating star.

\begin{figure}[htb]
\begin{center}
\includegraphics[width=75mm]{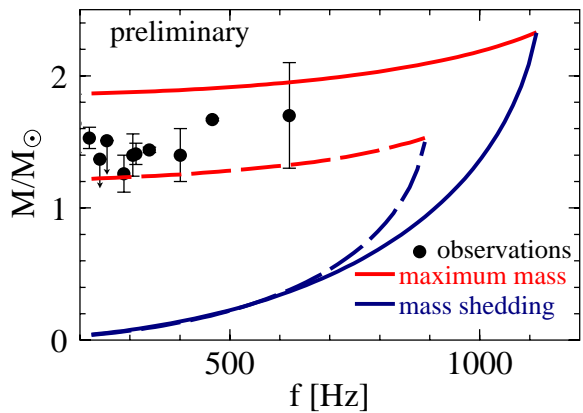}
\caption{ (Color online) Mass--frequency relation obtained with our models plotted against the observational data listed in \cite{kurk}. The solid and dashed curves represent the results obtained with and without screening, respectively.}
\label{M-f}
\end{center}
\end{figure}

Figure\ \ref{M-f} shows the result for a rotating star obtained using our EOSs with and without screening. The red curve shows the maximum mass of the star and the blue curve shows the mass-shedding curve, which corresponds to the Kepler frequency. The Kepler frequency indicates that the centrifugal force is equal in magnitude to gravity. Therefore, the area on the right-hand side of the blue curve is physically invalid. If the red curve is lower than the observations, the EOS should be ruled out. Our EOS in the screening case is thus consistent with these observations. However, our EOS without screening is not consistent and therefore inappropriate. This could be due to the softness of the EOS \cite{kurk}, although further studies are required.

\begin{figure}[htb]
\begin{center}
\includegraphics[width=75mm]{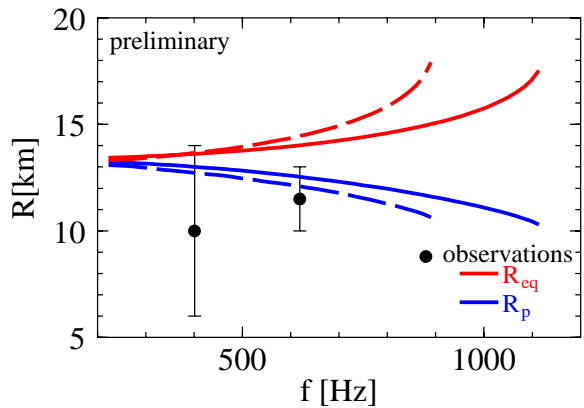}
\caption{ (Color online) Radius--frequency relation of our model plotted against observational data (SAXJ1808.4-3658 and 4U1608-52). The solid and dashed curves represent the results obtained with and without screening, respectively.}
\label{R-f}
\end{center}
\end{figure}

Figure\ \ref{R-f} suggests an important relation between the radius and rotation. The radius of a star is considered a single value because we ordinarily consider a star approximately spherical. However, if the star is rapidly rotating, it is an ellipse rather than a sphere, and we have to recognize the different radii.
Therefore, we introduce two values, $R_{\mathrm eq}$ and $R_{\mathrm p}$, which are the equatorial radius and polar radius, respectively. Figure\ \ref{R-f} shows $R_{\mathrm eq}$ and $R_{\mathrm p}$ with respect to rotation. If the rotation rate is 400 Hz or higher, the two radii are different. We thus have to note the effects of rotation on rapidly rotating stars.

\section{Summary and Concluding Remarks} \indent

We presented the difference between EOSs with and without charge screening taking into account rotation effects.
We used a simple model for quark matter and hadron matter.
To obtain a more realistic picture of the hadron--quark phase transition, we need to
take into account color superconductivity \cite{alf2,alf1,alf3} and relativistic mean field theory \cite{shen}. 
 A neutron star has another interesting feature---its magnetic field. The origin of the magnetic field is still unknown. A magnetic field can be explained by the spin-polarization of quark matter \cite{tat3,tat4}, but whether quark matter exists strongly depends on the EOS.
In this paper, we did not include magnetic fields. Several recent studies have investigated the effect of the magnetic field on the EOS \cite{dexh,chir,agui,mall}. Interesting results would be obtained if we took into account both the magnetic field and rotation effects.

\section*{Acknowledgments} \indent

This work was supported in part by a Principal Grant from the National Institute of Technology, Kagawa College. 
 

\begin{footnotesize}




\begin{thebibliography}{99}

\bibitem{rhic} K.~Adcox {\it et~al.}, (PHENIX collaboration),
	Phys. Rev. Lett. {\bf 88} 022301 (2002); C.~Adler {\it et~al.}, (STAR
	collaboration), Phys. Rev. Lett. {\bf 90} 082302 (2003).

\bibitem{mad3} J.~Madsen, Lect. Notes Phys. {\bf 516} 162 (1999).

\bibitem{chen} K.~S.~Cheng, Z.~G.~Dai and T.~Lu, Int. Mod. Phys. {\bf D7}
	139 (1998).

\bibitem{rev} For review, D.~H.~Rischke,
	Prog. Part. Nucl. Phys. {\bf 52} (2004) 197; J.~Macher and J.~Schaffner-Bielich, Eur. J. Phys. {\bf 26}
	341 (2005) and references therein.

\bibitem{pisa} R.~D.~Pisalski and F.~Wilczek,
	Phys. Rev. Lett. {\bf 29} 338 (1984).

\bibitem{latt} R.~V.~Gavai, J.~Potvin and S.~Sanielevici,
	Phys. Rev. Lett. {\bf 58} 2519 (1987).

\bibitem{end2}  T.~Endo, T.~Maruyama, S.~Chiba and T.~Tatsumi,
	Prog. Theor. Phys. {\bf 115} 337 (2006); hep-ph/0510279.

\bibitem{end3}  T.~Endo, Phys. Rev. {\bf C83} 068801 (2011).

\bibitem{end4}  T.~Endo, arXiv:1310.0913[astro-ph.HE].

\bibitem{kurk} A.~Kurkela, P.~Romatschke, A.~Vuorinen and B.~Wu, arXiv:1006.4062[astro-ph.HE].

\bibitem{orsa} M.~Orsaria, H.~Rodrigues, F.~Weber and G.A.~Contrera, Phys. Rev.  {\bf D87} 023001 (2013).

\bibitem{webe} F.~Weber, M.~Orsaria and R.~Negreiros, arXiv:1307.1103[astro-ph.SR]

\bibitem{belv} R.~Belvedere, K.~Boshkayev, J.~A.~Rueda, and R.~Ruffini, arXiv:1307.2836[astro-ph.SR]

\bibitem{dexh} V.~Dexheimer, R.~Negreiros and S.~Schramm, arXiv:1108.4479[astro-ph.HE].

\bibitem{chir} C.~Chirenti, and J.~Skakalala, Phys. Rev. {\bf D88} 104018 (2013).

\bibitem{agui} R.~Aguirre, E.~Bauer and I.~Vidana, Phys. Rev. {\bf C89} 035809 (2014).

\bibitem{mall} R.~Mallick and S.~Schramm, Phys. Rev. {\bf C89} 045805 (2014).

\bibitem{pet} H.~Heiselberg, C.~J.~Pethick and E.~F.~Staubo, Phys. Rev. Lett. {\bf 70} 1355 (1993).

\bibitem{gle2} N.~K.~Glendenning and S.~Pei, Phys. Rev. {\bf C52} 2250 (1995).

\bibitem{alf2} M.~Alford, K.~Rajagopal, S.~Reddy, and F.~Wilczek, Phys. Rev. {\bf D64} 074017 (2001).

\bibitem{maru}  Toshiki~Maruyama, T.~Tatsumi, D.N.~Voskresensky,
	T.~Tanigawa, T.~Endo and S. Chiba, Phys. Rev. {\bf C73} 035802 (2006); nucl-th/0505063.

\bibitem{maru2}  T.~Maruyama, T.~Tatsumi, T.~Endo and S.~Chiba, Recent Res. Devel. Phys. {\bf 7} 1 (2006); nucl-th/0605075

\bibitem{ster} N.~Stergioulas and J.L.~Friedman, Astrophys. J. {\bf 444} 306 (1995). 

\bibitem{alf1} For reviews, M.~Alford, A.~Schmitt,
	K.~Rajagopal and T.~Sch\"{a}fer, Rev. Mord. Phys. {\bf 80} 1455 (2008).

\bibitem{alf3} M.~Alford and S.~Reddy, Phys. Rev. {\bf D67} 074024 (2003).

\bibitem{shen} H.~Shen, H.~Toki, K.~Oyamatsu and K.~Sumiyoshi, Nucl. Phys. {\bf A637} 435 (1998). 

\bibitem{tat3} T.~Tatsumi, Phys. Lett. {\bf B489} 280 (2000).

\bibitem{tat4} T.~Tatsumi, arXiv:1107.0807[hep-ph].



\end{thebibliography}
%

\end{footnotesize}


\end{document}